\pdfoutput=1
\documentclass[%
 aip,
 jmp,%
 amsmath,amssymb,
 reprint,%
]{revtex4-1}
\usepackage{graphicx}
\usepackage{dcolumn}
\usepackage{amsmath}
\usepackage{bm}
\usepackage{textcomp}
\usepackage{pbox}
\usepackage{array,booktabs}

\begin{document}

\preprint{AIP/123-QED}

\title[]{High Resolution Angle Resolved Photoemission with Tabletop 11eV Laser}

\author{Yu He}
\affiliation{
SIMES, SLAC National Accelerator Laboratory, Menlo Park, California 94025
}
\affiliation{
Department of Applied Physics, Stanford University, Stanford, California 94305
}
\author{Inna M. Vishik}
\affiliation{
SIMES, SLAC National Accelerator Laboratory, Menlo Park, California 94025
}
\affiliation{
Department of Applied Physics, Stanford University, Stanford, California 94305
}
\author{Ming Yi}
\affiliation{
SIMES, SLAC National Accelerator Laboratory, Menlo Park, California 94025
}
\affiliation{
Department of Applied Physics, Stanford University, Stanford, California 94305
}
\author{Shuolong Yang}
\affiliation{
SIMES, SLAC National Accelerator Laboratory, Menlo Park, California 94025
}
\affiliation{
Department of Applied Physics, Stanford University, Stanford, California 94305
}
\author{Zhongkai Liu}
\affiliation{
SIMES, SLAC National Accelerator Laboratory, Menlo Park, California 94025
}
\affiliation{
Department of Physics, Stanford University, Stanford, California 94305
}
\author{James J. Lee}
\affiliation{
SIMES, SLAC National Accelerator Laboratory, Menlo Park, California 94025
}
\affiliation{
Department of Applied Physics, Stanford University, Stanford, California 94305
}
\author{Sudi Chen}
\affiliation{
SIMES, SLAC National Accelerator Laboratory, Menlo Park, California 94025
}
\affiliation{
Department of Applied Physics, Stanford University, Stanford, California 94305
}
\author{Slavko Rebec}
\affiliation{
SIMES, SLAC National Accelerator Laboratory, Menlo Park, California 94025
}
\affiliation{
Department of Applied Physics, Stanford University, Stanford, California 94305
}
\author{Dominik Leuenberger}
\affiliation{
SIMES, SLAC National Accelerator Laboratory, Menlo Park, California 94025
}
\affiliation{
Department of Applied Physics, Stanford University, Stanford, California 94305
}
\author{Alfred Zong}
\affiliation{
Department of Physics, Stanford University, Stanford, California 94305
}

\author{C. Michael Jefferson}
\affiliation{
Lumeras LLC, 207 McPherson St, Santa Cruz, California 95060
}
\author{Robert G. Moore}
\author{Patrick S. Kirchmann}
\affiliation{
SIMES, SLAC National Accelerator Laboratory, Menlo Park, California 94025
}
\author{Andrew J. Merriam}
\affiliation{
Lumeras LLC, 207 McPherson St, Santa Cruz, California 95060
}
\author{Zhi-Xun Shen}
\affiliation{
SIMES, SLAC National Accelerator Laboratory, Menlo Park, California 94025
}
\affiliation{
Department of Applied Physics, Stanford University, Stanford, California 94305
}
\email{zxshen@stanford.edu}

\date{\today}

\begin{abstract}
We developed a table-top vacuum ultraviolet (VUV) laser with $113.778$~nm wavelength (10.897eV) and demonstrated its viability as a photon source for high resolution angle-resolved photoemission spectroscopy (ARPES). This sub-nanosecond pulsed VUV laser operates at a repetition rate of 10~MHz, provides a flux of 2$\times$10$^{12}$ photons/second, and enables photoemission with energy and momentum resolutions better than 2~meV and 0.012~\AA$^{-1}$, respectively. Space-charge induced energy shifts and spectral broadenings can be reduced below 2~meV. The setup reaches electron momenta up to 1.2~\AA$^{-1}$, granting full access to the first Brillouin zone of most materials. Control over the linear polarization, repetition rate, and photon flux of the VUV source facilitates ARPES investigations of a broad range of quantum materials, bridging the application gap between contemporary low energy laser-based ARPES and synchrotron-based ARPES. We describe the principles and operational characteristics of this source, and showcase its performance for rare earth metal tritellurides, high temperature cuprate superconductors and iron-based superconductors.
\end{abstract}

\maketitle


\section{\label{sec:level1}INTRODUCTION}

Angle-resolved photoemission spectroscopy (ARPES) directly accesses electronic band structures and electronic self-energies in a momentum-resolved manner, which makes it a powerful and unique technique in condensed matter research. This momentum resolution allows one to distinguish multiple electronic band dispersions in multiband systems, to assess the symmetry of order parameters, and to characterize electronic anisotropy in solid state systems. Recent progress in the ARPES technique has pushed energy resolutions down to meV scales, which gives unprecedented access to low energy excitations in quantum materials. With its unique combination of energy and momentum resolution, ARPES contributes significantly to the understanding of many two-dimensional quantum materials \cite{Ohta2006,Zhou2007,Johnson2002133,Kordyuk2014}, in particular high T$_c$ superconducting cuprates \cite{Damascelli2003,Ding1996}, iron based superconductors \cite{Stewart2011,Donghui2012,JJ2014} and topological states of matter \cite{Hsieh2008,Chen2009}.

ARPES collects photoemitted electrons as a function of kinetic energy $E_k$ and emission angle $\theta$ with respect to the sample surface normal. The conservation of energy and in-plane momentum of each electron\footnote{In general, the out-of-plane momentum is not conserved as the electron overcomes the surface potential barrier during the photoemission process.} allows the calculation of the electron's binding energy with respect to the Fermi level $E_F$ and its parallel momentum $\hbar k_{\parallel}$:\cite{Damascelli2003}

\begin{equation}
E - E_F = E_k + \phi - h\nu
\tag{1}\label{eq:1}
\end{equation}
\begin{equation}
\hbar k_{\parallel} = \sqrt{2m_eE_k} ~sin\theta 
\tag{2}\label{eq:2}
\end{equation}

Here, $\phi$ denotes the work function of the material and $h\nu$ the photon energy. Eq.~(\ref{eq:1}) describes how larger photon energies provide access to states with higher binding energy. Eq.~(\ref{eq:2}) shows that the accessible electron momenta are limited by the electron kinetic energy, which scales monotonically with the photon energy. For typical values of $\theta$ = 50$^{\circ}$ and $\phi$ = 4.5~eV, the minimum photon energy required to capture the entire first Brillouin zone (BZ) of a material with 3.5~\AA~lattice constant is $h\nu$ = 9.7~eV.\footnote{$\theta$ = 35.0$^{\circ}$ (maximum sample surface rotation) + 15.0$^{\circ}$ (typical detector acceptance angle)}

The relation between photon energy $h\nu$ and parallel momentum resolution $\Delta k_{\parallel}$ is given by
\begin{equation}
\hbar \Delta k_{\parallel}(E = E_F) = \sqrt{2m_e(h\nu - \phi)} ~cos\theta \Delta \theta
\tag{3}\label{eq:3}
\end{equation}

As indicated in Eq.~(\ref{eq:2}-\ref{eq:3}), for a given detector angular resolution $\Delta \theta$ and momentum feature, higher photon energies degrade the momentum resolution due to increasing $h\nu$ and decreasing $\theta$.


When employing pulsed photon sources Coulomb repulsion of electrons photoemitted by a single pulse can lead to space charging effects, which both shift and broaden the spectrum.\cite{Graf2010,Passlack2006} The effects of space charging can be mitigated by higher repetition rates while maintaining a constant photon flux. Therefore to achieve ultimate resolution and applicability, a light source optimized for high-resolution ARPES studies of complex materials should satisfy several requirements.
\begin{enumerate}
    \item Sufficiently high photon energy ($>$~10~eV) to access the first BZ and probe valence bands (Eq.~(\ref{eq:1}-\ref{eq:2}))
    \item Sufficiently low photon energy ($<$~20~eV) to facilitate high momentum resolution (Eq.~(\ref{eq:3}))
    \item Sufficiently high repetition rate ($\gg$~100~kHz) to limit the loss of energy and momentum resolutions due to space-charge effects while maintaining sufficiently high signal-to-noise ratio
    \item Sub-meV bandwidth, high flux and long term stability
    \item Variable polarization for control over photoemission matrix elements
    \item Additional requirements may include a small ($<$~1~mm) beam spot and short ($\leq$~100~ps) pulse duration for time-of-flight detection schemes
\end{enumerate}

While it is challenging to integrate all these properties into one single light source, existing light sources successfully capture different aspects of the requirements listed above (TABLE I). Noble gas discharge lamps, helium and xenon in particular, were among the earliest photoemission light sources with photon energies in the range of 8.4 - 41~eV.\cite{Harter2012,zhang2012photoemission,Souma2007,VGVUV5000} These pioneering experiments established ARPES as a valuable tool for the analysis of electronic structures in solids.

Continuous advances of synchrotron technology provided bright and tunable light sources that were pivotal to the tremendous success of ARPES in the last two decades. Comparing with gas discharge lamps, synchrotron-based ARPES features small ($\textless$ 0.5~mm) beam spots which improve momentum resolution, and avoids reduced sample lifetimes due to gas molecules effusing from the light source.\cite{Harter2012,zhang2012photoemission} Additionally, third generation synchrotron sources feature low photon energies in combination with excellent energy-momentum resolution.\cite{Strocov2010,Strocov2014} Despite the outstanding capabilities of synchrotron-based ARPES the demand for accessible, table-top ultraviolet light sources with higher photon flux and improved energy stability remains high. 

Laser-based ARPES commonly utilizes non-linear optical crystals to up-convert infrared light pulses into the UV spectral region.\cite{Wu1996} The generated low photon energies of 6-7~eV result in the excellent energy and momentum resolution required for studies of detailed band dispersions and subtle low-energy excitations.\cite{Kiss2008,Kirchmann2008,Rui2014,Damm2015} Yet these low photon energies limit the access to high momenta and valence bands.

Although highly desirable, frequency up-conversion to sub-170 nm wavelengths is not feasible in nonlinear optical crystals due to the re-absorption in the vacuum UV (VUV) range in any material. Instead, these short wavelengths can be generated via high-harmonic generation in polarizable gases. Tjernberg and coworkers built a 10.5~eV light source dedicated to ARPES by frequency tripling the third harmonic of a pulsed IR laser in Xe gas,\cite{Tjernberg2011} which, when integrated with a time-of-flight detector, for the first time made laser-based photoemission study possible at many materials' BZ boundaries. Such non-resonant XUV generation schemes usually require MW peak power, which in turn typically limit the laser repetition rate to $\textless$ 1~MHz.\cite{Tjernberg2011} The capabilities of commonly used light sources for ARPES are summarized in TABLE I.

\begin{table*}[t]
\resizebox{\textwidth}{!}{%
\centering
\setlength\extrarowheight{15pt}
\begin{tabular}{cccccccc}

\hline
 & \pbox{20cm}{$\Delta$E \\ (meV)} & \pbox{20cm}{spot FWHM \\ (mm)} & \pbox{20mm} {max k$_\parallel$\\ (\AA $^{-1}$)} & \pbox{20cm} {max BE \\ $E - E_F$ (eV)} &  \pbox{20cm} {polarization \\ control} & \pbox{20cm}{rep rate \\ (MHz)} & \pbox{20cm} {photon flux \\ (photons/second)} \\ \hline

5.8 - 8eV laser\cite{Koralek2007,Kiss2008,Kiss2005}  & $\textless$~2 & $\textless$~0.1 & $\textless$~0.75 & $\textless$~2 & Yes & $\sim$10$^2$ & $\sim$10$^{14}$ \\
He/Xe discharge lamp\cite{Harter2012,zhang2012photoemission,Souma2007} & $\textless$~10 & $\sim$~1 & \pbox{20cm} {$\textless$~0.82 \\ $\textless$~1.8 \\ $\textless$~2.7} & \pbox{20cm} {$\textless$~3 (XeI)\\ $\textless$~16 (HeI$\alpha$) \\ $\textless$~36 (HeII)} & No & CW & $\sim$~10$^{13}$\\
Synchrotron\cite{BL54,BL403,BESSY1cube} & $\textgreater$~1 & $\textless$~0.3 & multiple BZ & \pbox{20cm}{up to \\ core level\\ (keV)} & Limited & 500 & $\sim$~10$^{12}$\\ 
11eV laser & $\textless$~2 & $\textless$~0.5 & $\textless$~1.2 & $\textless$~6 & Yes & 1$\sim$20 & $\sim$10$^{13}$ \\ \hline

\end{tabular}}
\caption{Comparison of contemporary ARPES light source properties. 11eV laser combines high energy-momentum resolution and large energy-momentum coverage into one single light source. The maximum parallel momentum are calculated based on 4.5eV work function of typical materials.}
\end{table*}

In this article, we describe a table-top pulsed VUV light source optimized for high resolution ARPES studies of quantum materials. Its photon energy of $\sim$ 11~eV is sufficient to cover the complete BZ in most materials, while maintaining exceptional energy and momentum resolution at high photon flux. We demonstrate its capabilities for high resolution ARPES on a number of materials, including antimony, rare earth metal tritelluride, unconventional copper and iron-based high temperature superconductors and single unit cell FeSe film. 

\section{\label{sec:level1}SYSTEM OVERVIEW}
\subsection{\label{sec:level2}Vacuum-Ultraviolet (VUV) Laser Light Source}
 \begin{figure*}
 \includegraphics[width=0.9\textwidth]{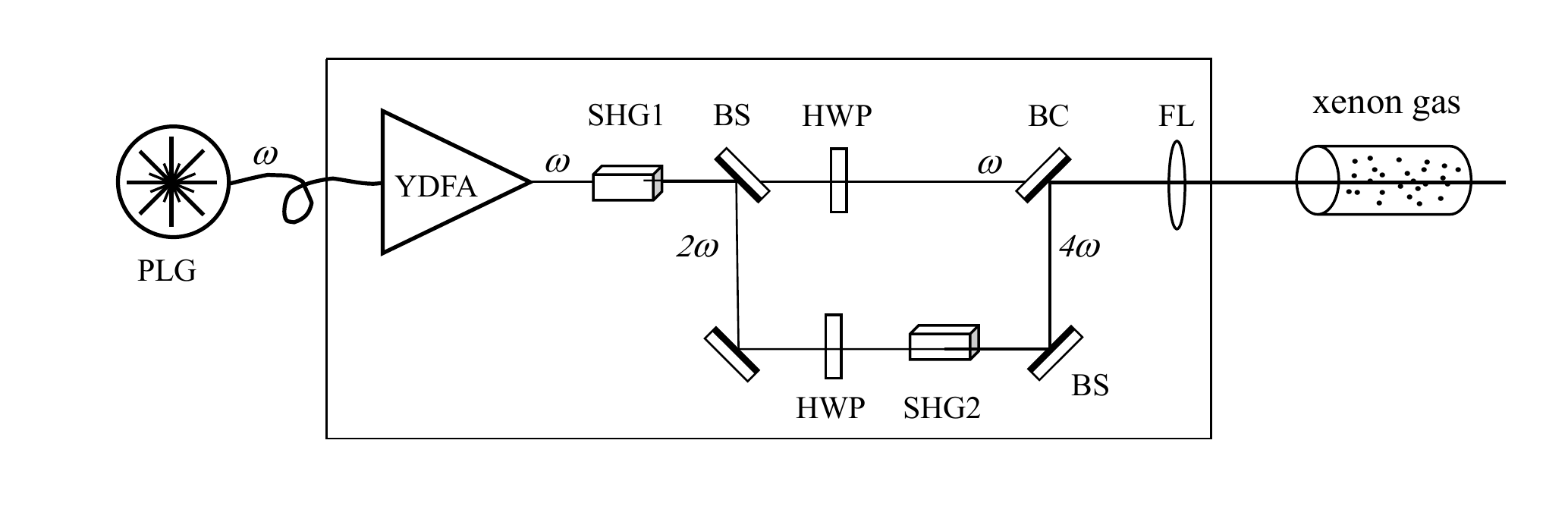}
 \caption{\label{Fig1}Schematic diagram of the coherent VUV light source. PLG, fiber-coupled Pulsed Light Generator seed source; YDFA, Yb-doped fiber amplifier; SHG, Second Harmonic Generation nonlinear crystal stages; HWP, half-wave plate; BS, Beam Splitter; BC, Beam Combiner; FL, Focus Lens. VUV light is produced by frequency-mixing the fundamental ($\omega$) and fourth-harmonic (4$\omega$) in an external xenon gas cell via the third-order nonlinear susceptibility $\chi^{(3)}$[9$\omega$; 4$\omega$ + 4$\omega$ + $\omega$]}
 \end{figure*}

Our coherent VUV light source utilizes three cascaded stages of nonlinear frequency conversion of a quasi-continuous wave pulsed, 1024~nm infrared (IR) solid-state laser. As shown schematically in Fig.~\ref{Fig1}, the first two conversion stages occur in birefringent nonlinear crystals; VUV flux is generated via two-photon resonant, sum-frequency generation in xenon gas using the fundamental ($\omega$) and fourth-harmonic (4$\omega$) of this laser system.

The VUV light source is driven by a fiber-coupled wavelength-tunable (1024~nm) IR seed source with variable repetition rate and sub-nanosecond pulse duration. Its output is amplified by a Yb-doped fiber amplifier to an average power of 10~W, and a peak power of $\sim$10~kW. The amplified IR light is up-converted to the second harmonic (512~nm) using a first second harmonics crystal (SHG1) with 50\% efficiency. The second harmonic light is separated from the fundamental using a dichroic beam splitter (BS) and refocused into a second nonlinear crystal (SHG2) to generate UV light (256~nm). By controlling the polarization of the 512~nm beam, the output power of the frequency quadrupled UV light can be continuously modulated. This attenuation method provides VUV (114~nm) flux control, without change to the alignment or beam profile.

\begin{figure}
 \includegraphics[width=0.35\textwidth]{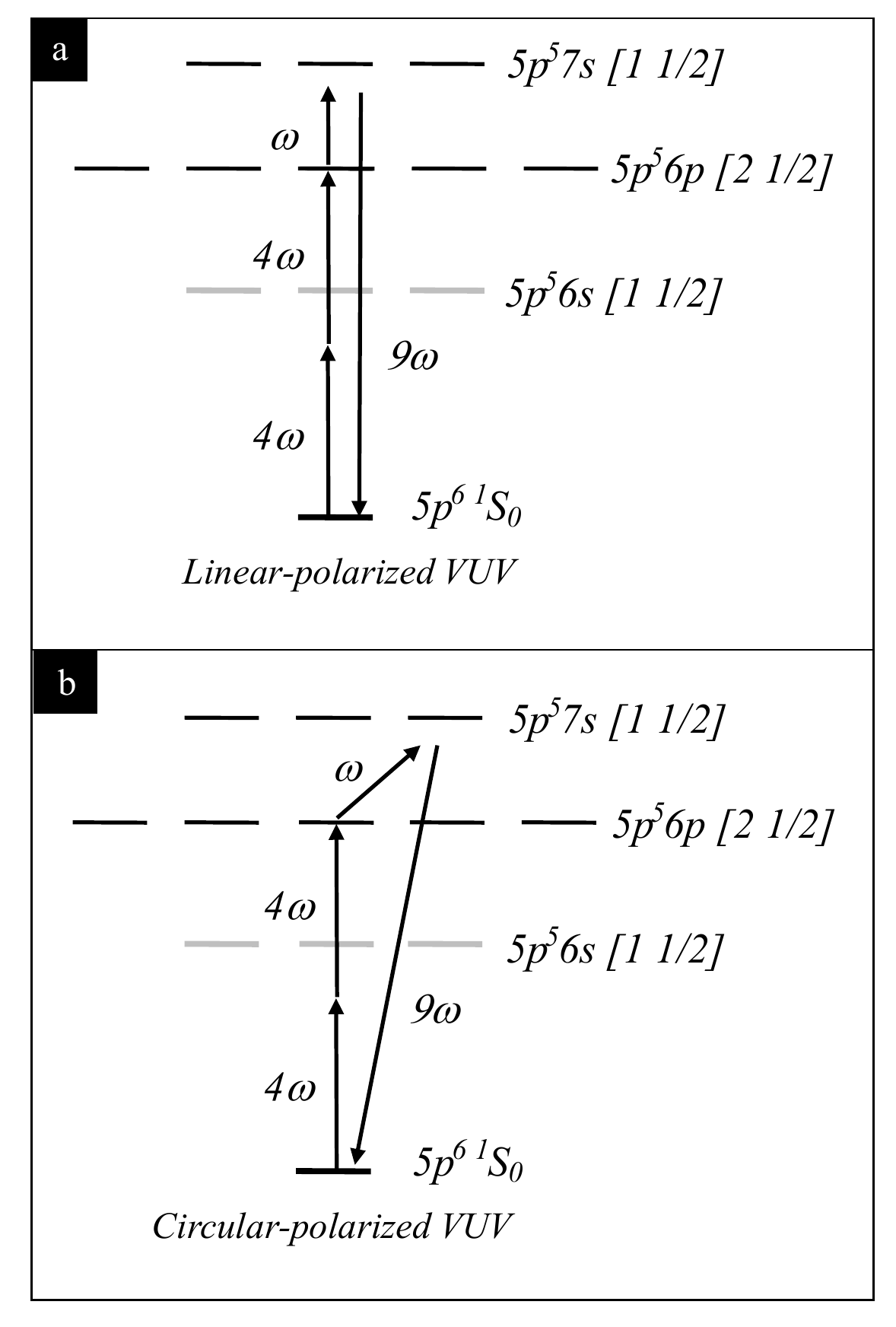}
 \caption{\label{Fig2} Energy-level diagram of the nonlinear process in xenon gas, demonstrating polarization control of the upconverted VUV photon.  Two fourth-harmonic UV photons drive the $5p^6 ~^1S_0$ - $5p^5 6p$ two-photon transition at resonance. This local atomic oscillator beats against the applied fundamental wave to generate sum frequency (9$\omega$). Optical polarizations are indicated by the direction of the arrows: vertical arrows correspond to linear polarization (zero change in angular momentum) and angled arrows correspond to circular (angular momentum changes by $\pm \hbar$). When the polarization of the UV light (4$\omega$) is linear, the polarization of the VUV light (9$\omega$) may be adjusted from linear (a), to circular (b), depending on the polarization of the fundamental IR ($\omega$).}
 \end{figure}

Following the beam splitter, the polarization of the residual IR beam is controlled by a second half-wave plate (HWP) to achieve any given VUV beam polarization. Then the IR and UV beams are overlapped in time and space using dichroic dielectric-coated beam-combiner mirrors, and focused by a single lens into a xenon-filled gas cell. The energy-level diagram of the specific nonlinear process in atomic xenon, and the control of the final VUV polarization, are shown in Fig.~\ref{Fig2}. Two UV photons (4$\omega$) with wavelengths of 256.015~nm resonantly drive the dipole forbidden $5p^6 ~^1S_0$ - $5p^5 6p$ transition.\cite{Moore1953,Gornik1981,Plimmer1989} By mixing this local atomic oscillator with a fundamental IR photon ($\omega$), light at the 9$^{th}$ harmonic (9$\omega$) is generated (10.897~eV, or 113.778~nm). By driving the two-photon transition with linearly-polarized UV light, the polarization of the VUV photon is determined by the optical polarization of the IR photon.

Traditional gas-phase nonlinear optical systems have typically required MW-scale peak-power driving lasers for efficient conversion, thereby severely limiting the achievable repetition rates. However, by operating at a two-photon resonance condition, the peak optical powers required for efficient up-conversion are reduced to the kW level, which is a necessary condition for increasing the repetition rate of the source to the MHz range,\cite{Tjernberg2011} and for reducing the overall source size to a table-top device.

 \begin{figure}[h]
 \includegraphics[width=0.35\textwidth]{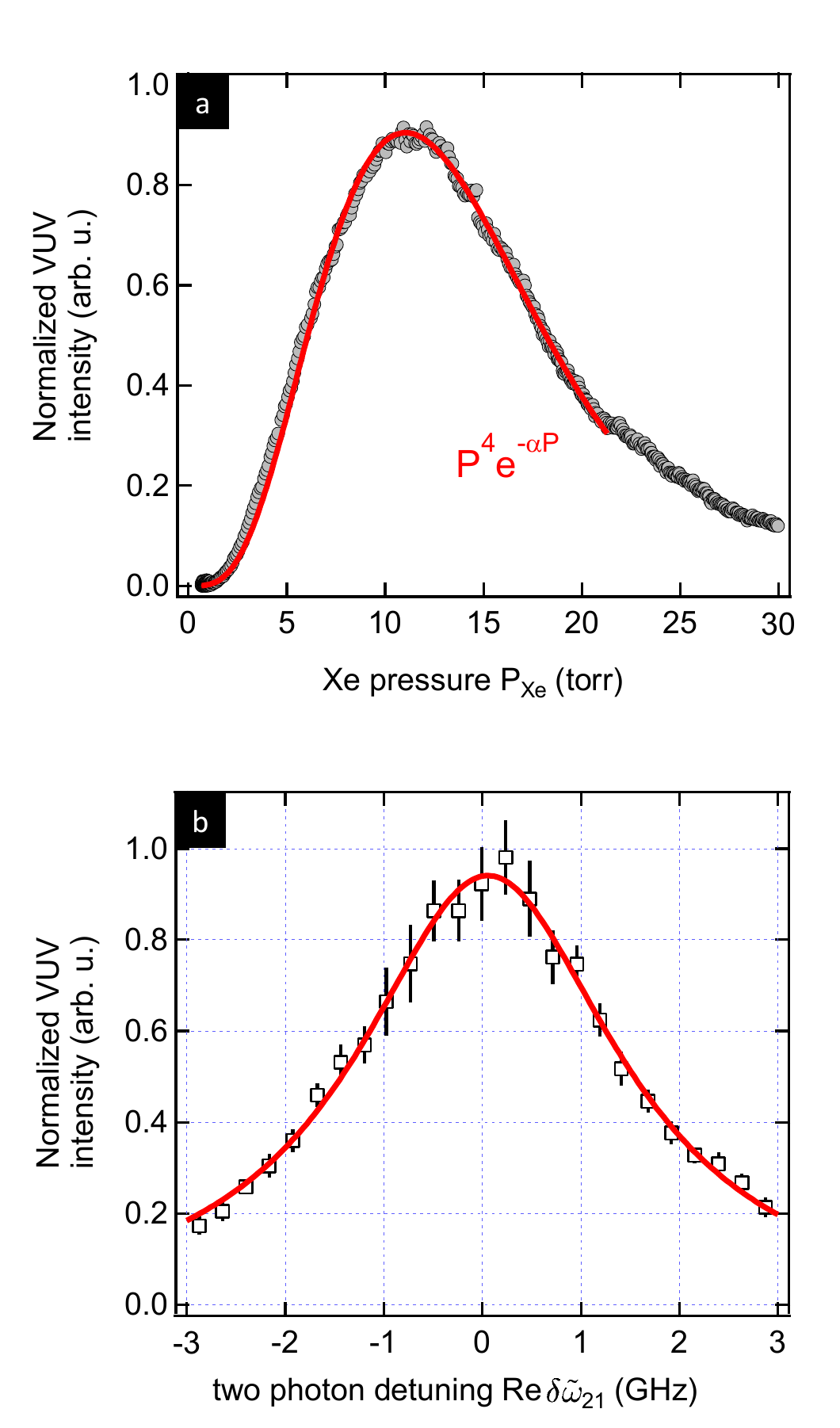}
 \caption{\label{Fig3} (a) Dependence of 11~eV flux on xenon pressure. The data are fit to a tight-focussing model,\cite{Vidal1980} indicating excellent beam quality and minimal saturation. (b) Resonant enhancement of the VUV up-conversion process at 1~ns pulsewidth and 1~MHz repetition rate. The generated VUV flux scales as Im($1/\delta \tilde{\omega}$), with the complex detuning parameter $\delta \tilde{\omega}$ defined as ($\omega_{21}$ $-$ $8\omega$) + i$\Gamma$; $\omega_{21}$ is the energy of the $5p^6 ~^1S_0$ - $5p^5 6p$ two-photon transition (9.69~eV), $\omega$ is the frequency of the fundmental IR light, and $\Gamma$ is the combination of the transition linewidth and laser linewidth $\sim$ 1~GHz.}
 \end{figure}

Atomic xenon is negatively dispersive for degenerate two-photon-resonant sum-frequency generation, which enables a tightly-focussed geometry for VUV generation.\cite{Vidal1980} Fig.~\ref{Fig3}(a) demonstrates typical variation of generated VUV flux on the xenon gas pressure. The single-peaked curve, with a maximum at approximately 12 torr xenon pressure, is typical of tightly-focussed sum-frequency generation.\cite{Bjorklund1975}

Due to the intermediate resonance step in the frequency-conversion process, the generated VUV flux is particularly sensitive to the absolute frequency of the UV beam. As shown in Fig.~\ref{Fig3}(b) for the 1~MHz repetition rate, 1~ns pulsewidth laser setup, detuning of the UV wave from the two-photon resonance results in a rapid reduction in VUV generation efficiency.\cite{Harris1976} The width of this resonance profile is set by the convolution of the laser linewidth and the natural linewidth (including isotopic and Doppler broadening) of the two-photon transition. In order to maintain peak flux, the central frequency of the UV beam must be maintained to within 1~GHz; the IR frequency must therefore be stabilized to the central frequency within 250~MHz uncertainty. In comparison, the total width of the resonance curve for the case of the broader-bandwidth excitation laser pulses produced in the 10~MHz, 100~ps laser configuration reaches $\sim$8~GHz. This provides a lower bound on both the energy resolution and the absolute energy stability of the 11~eV laser of $\sim$ 30~$\mu$eV.

\subsection{\label{sec:level2}VUV DISPERSION AND FOCUSING}

 \begin{figure*}[t]
 \includegraphics[width=0.9\textwidth]{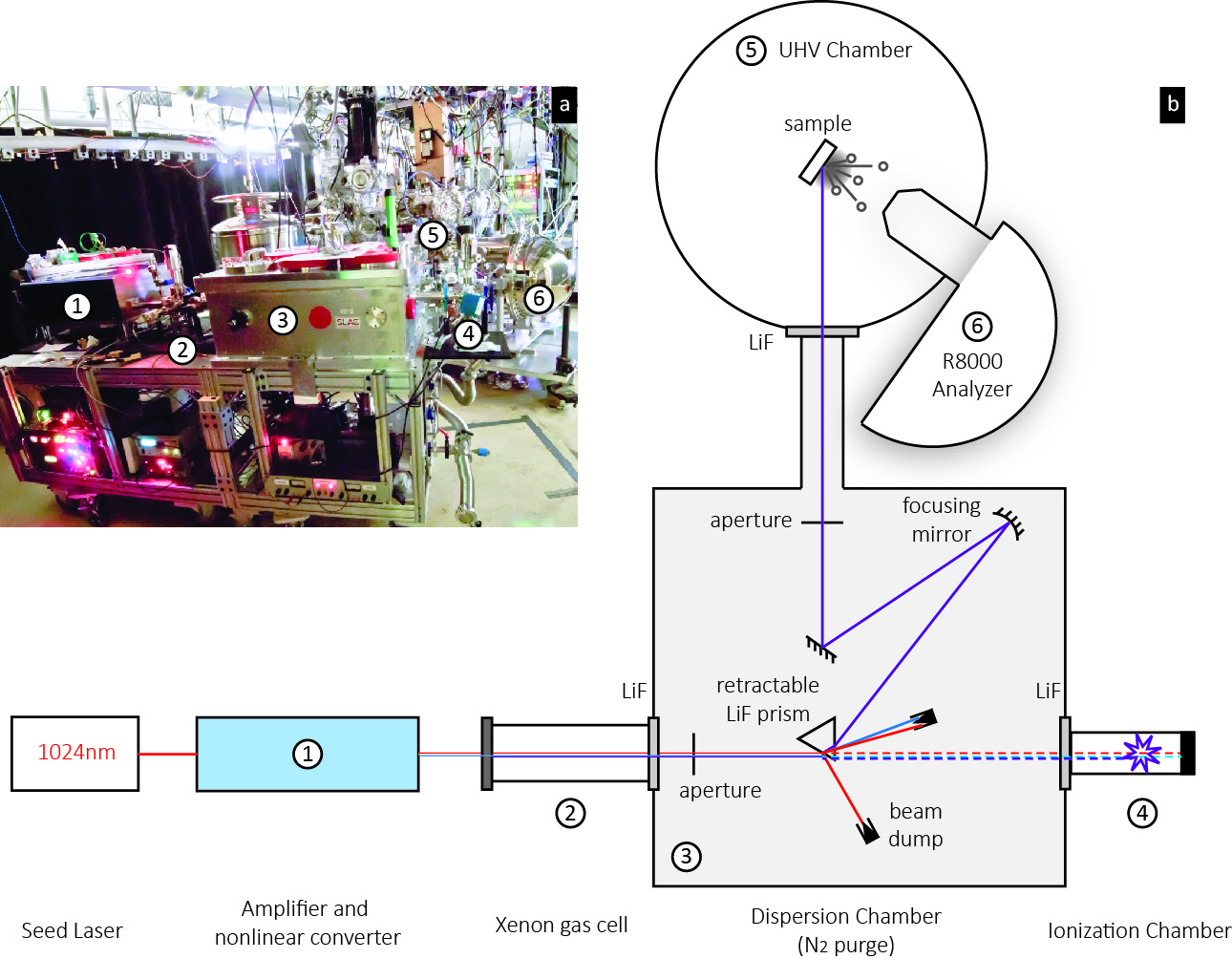}
 \caption{\label{Fig4}(a) Picture of the experimental setup (b) Schematic layout of the optics setup and its coupling to the ARPES UHV chamber.}
 \end{figure*}

VUV light is absorbed strongly by several atmospheric constituents, most notably oxygen and water vapor.\cite{Krupenie1972,Tennyson2009} The 1/e attenuation length for 11~eV photons in air is approximately 3~mm.\cite{Watanabe1953} Therefore, the entire beam path for the 11~eV light must be either evacuated or purged to remove the absorbing species.

The xenon-filled gas converter is connected directly to a N$_2$-purged dispersion chamber using an uncoated LiF window. The 11~eV light is generated inside the xenon gas cell as a nearly-diffraction-limited beam that propagates collinearly with the driving IR and UV beams. The IR and UV photons have energies of 1.21~eV and 4.84~eV, respectively, and Watt-level average powers. Illumination of the sample with these IR and UV beams can cause heating and photoemission and must be avoided. We employ an equilateral LiF prism to refractively disperse the 11~eV flux from the lower-energy photons prior to photoemission. The prism is operated in a minimum-deviation condition, so that the incidence angle is near the Brewster angle for the VUV light, thus minimizing reflective losses for horizontal-linearly polarized 11~eV flux.\cite{Laporte1982}

Under normal photoemission measurement conditions, the prism is positioned to refract the co-propagating beams; both the front-surface reflections, and the refracted IR and UV beams, are trapped by beam dumps to minimize scattered light. In order to measure the 11~eV source power, the prism may be automatically retracted to allow the three laser beams to propagate to the opposite port of the dispersion chamber (dashed line in Fig.~\ref{Fig4}(b)). The photocurrent produced by an ionization chamber mounted to this port (Lumeras model IC-LF-C, with a LiF entrance window and filled with isopropanol vapor) provides solar-blind measurements of the 11~eV flux.

The 11~eV light is then focused onto the ARPES sample using a single concave reflective MgF$_2$-overcoated aluminum mirror. This mirror images the laser waists in the xenon converter onto the ARPES sample with a magnification factor of approximately 3. The 11~eV photons propagate through a second, uncoated 2¬mm-thick LiF window into the UHV chamber. Photoelectrons are collected by a Scienta R8000 hemispherical electron analyzer.

\section{\label{sec:level1}SYSTEM CHARACTERIZATION AND BENCHMARK}
\subsection{\label{sec:level2}Ultimate resolution}

 \begin{figure}
 \includegraphics[width=0.45\textwidth]{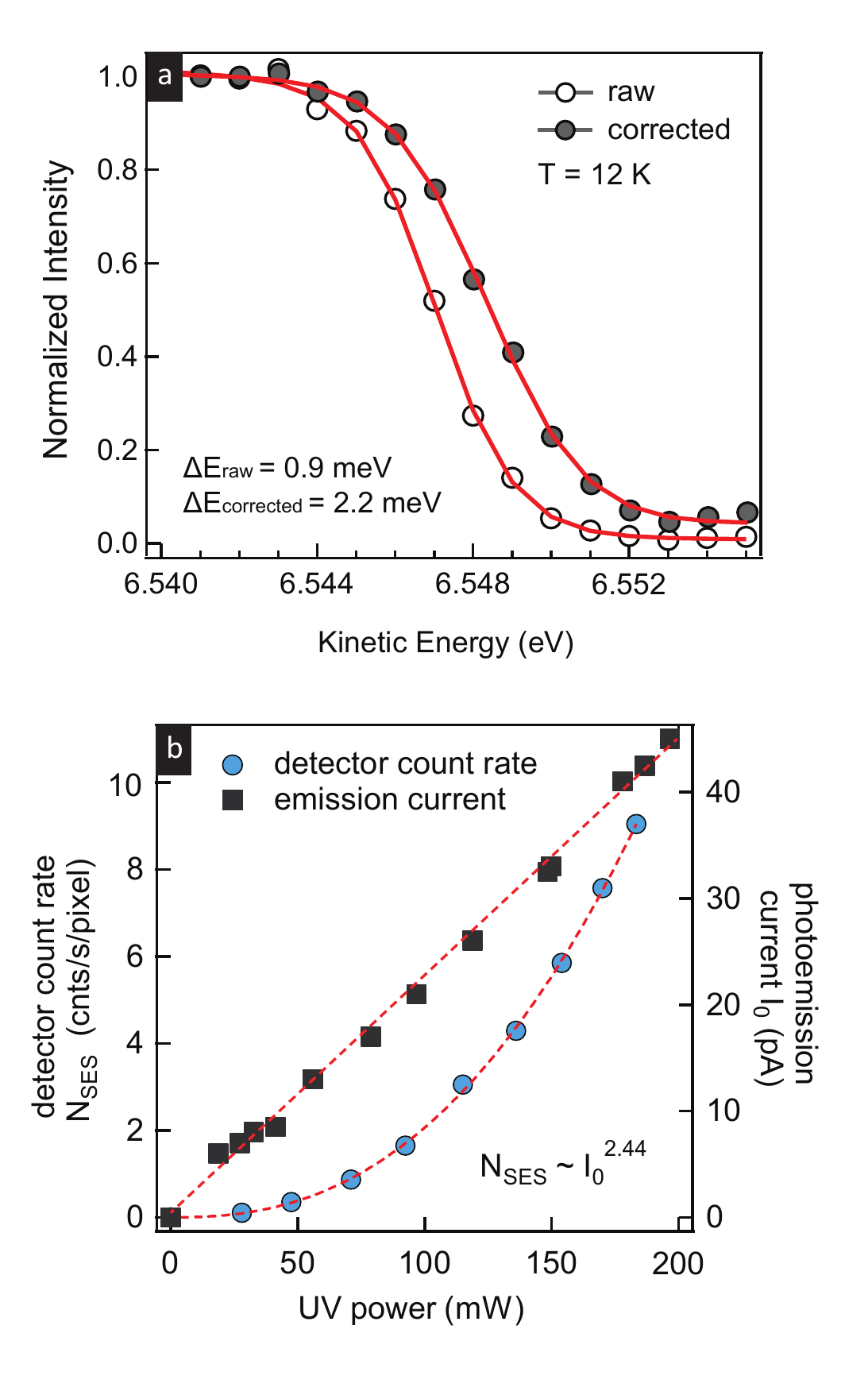}
 \caption{\label{Fig5} (a) Characterization of the energy resolution. Angle integrated energy distribution curve (EDC) of an evaporated polycrystalline Au film at 12K. The raw curve has an artificial sharpening and downshifting of the Fermi edge due to detector count-rate nonlinearity. This effect is removed in the corrected curve. (b) Detector nonlinearity characterization with a 7~eV light source. Detector count rate is measured as a function of photoemission current, which can be described by a power law fitting.}
 \end{figure}

Fig.~\ref{Fig5}(a) shows the momentum-integrated spectrum of an evaporated polycrystalline gold film. The angle integrated Energy distribution curves (EDC) was fit with a Fermi-Dirac function convolved with a Gaussian instrument energy-resolution function (red lines), giving 0.9~meV as an upper bound of the experimental energy resolution for the raw spectrum (open markers). A similar analysis was carried out after correcting the spectrum for detector nonlinearity (solid markers). The correction gives an experimental resolution of 2.2~meV. The origin and correction procedure of the detector nonlinear effect were detailed in Ref[40]. As shown in Fig.~\ref{Fig5}(b), by continuously varying the laser power, the detector count rate was measured as a function of total photoemission current over a large photon flux range. To remove the nonlinearity, detector count rates were mapped to the corresponding photoemission current (incident laser power) for every pixel.\cite{Reber2014}

\begin{figure}
 \includegraphics[width=0.45\textwidth]{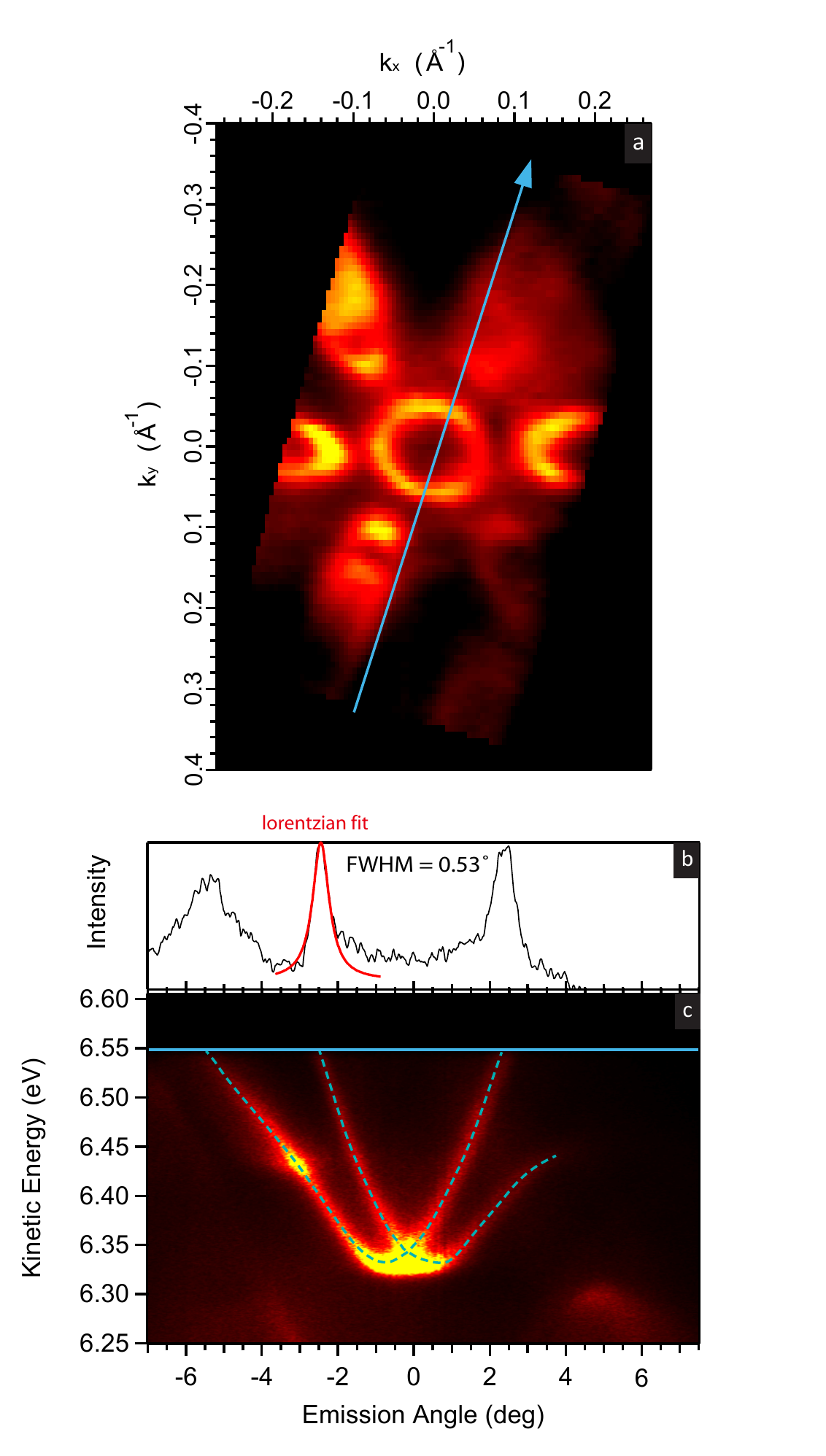}
 \caption{\label{Fig6} Characterization of the momentum resolution. (a) Fermi surface of the Sb(111) surface state. The blue line cut is shown in (c), where dashed lines are rashba-split surface states.\cite{Takayama2014} (b) angular distribution curve at $E_F$}
 \end{figure}

Antimony is known to host a metallic surface state on the (111) surface,\cite{Sugawara2006} which we can use to estimate the upper bound of the setup\rq s momentum resolution as shown in Fig.~\ref{Fig6}. A bulk Sb crystal is cleaved \emph{in-vacuo} on the (111) surface with Fig.~\ref{Fig6}(c) showing the band dispersion along the $\Gamma$-M cut. The angular distribution curve of the surface state from inner $\Gamma$ pocket at $E_F$ reaches a FWHM of 0.53$^{\circ}$, or 0.012 \AA$^{-1}$ when converted to parallel momentum. It should be noted that this is a combined effect from both instrument resolution and intrinsic sample-dependent electron scattering at $E_F$. In comparison, high quality measurements on optimally doped Bi-2212 can yield 0.37$^{\circ}$ angular or 0.0039 \AA$^{-1}$ momentum FWHM at $E_F$ with 7~eV laser-ARPES.\cite{Ishizaka2008}

\subsection{\label{sec:level2}Space charging}

\begin{figure*}[t]
 \includegraphics[width=0.9\textwidth]{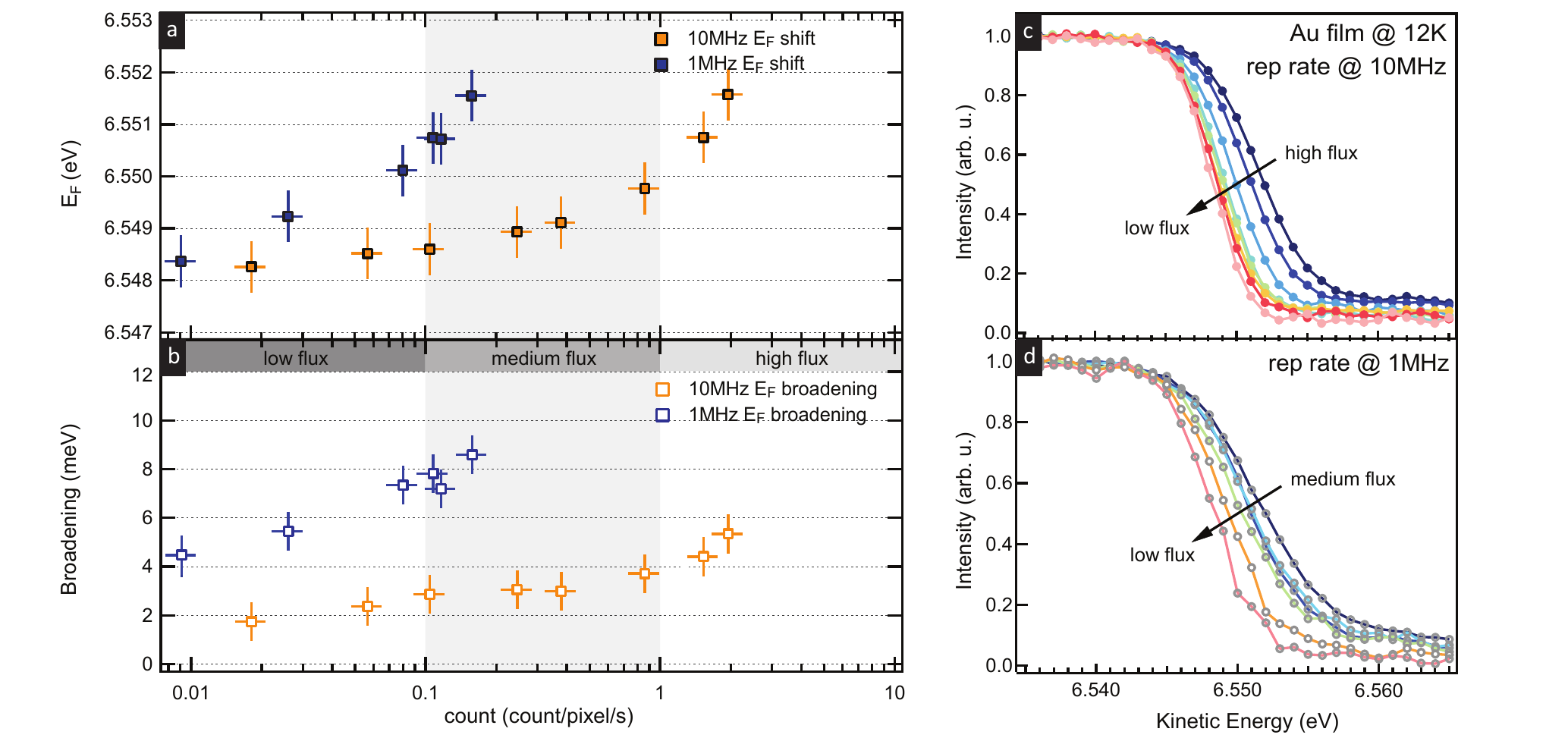}
 \caption{\label{Fig7} Characterization of space charging effects. (a) Flux dependent Fermi level shift. (b) Flux dependent broadening of Fermi edge (combined instrument resolution). Orange markers are data taken with 100~ps pulse duration and 10~MHz repetition rate, and the blue markers with 500~ps and 1~MHz. Flux dependent integrated EDC from polycrystalline gold were taken at 12K, at a repetition rate of 10~MHz (c) and 1~MHz (d) respectively. Our typical operating flux is between 0.1 and 1 count/pixel/second (medium flux range).}
  \end{figure*}

We employ two systems with comparable flux but different repetition rates to calibrate the system's space-charging effect. For our control experiment setup, at 1~MHz repetition rate, 500~ps pulse duration and $2\times10^{12}$ photons/second, space charging caused the measured Fermi level (on evaporated polycrystalline gold) to shift by 6~meV.

In order to mitigate the space charging induced spectral changes, we reduce the pulse duration to 100~ps while maintaining the peak pulse intensity of the IR beam, which is crucial to preserve high conversion efficiency of VUV photon generation.\cite{Hellmann2009} The repetition rate is boosted up to 10~MHz to compensate for the reduced photon count within every single pulse comparing to 1~MHz setup.

Fig.~\ref{Fig7} shows the photon-intensity dependence of space charging for both the 1~MHz and the 10~MHz setup. The severity of space charging is quantified both by the shift in $E_F$ (Fig.~\ref{Fig7}(a)) as well as the broadening of the Fermi edge (Fig.~\ref{Fig7}(b)). The broadening is defined as the fitted instrument resolution as is adopted in Fig.~\ref{Fig5}(a). Electron count/pixel/second from the Scienta R8000 detector camera is used as a measure of beam intensity. We find that up to 10$^{12}$ photons/sec or 1 electron count/pixel/sec, space charging can be limited to 2~meV $E_F$ shift and 4~meV broadening for the 10~MHz configuration. In the control experiment (blue markers), the 1~MHz setup gives worse space charging with a decade lower photon flux. In typical high resolution ARPES experiment operating at 0.1-1 count/pixel/sec, space charging effects are almost absent in the 10~MHz setup.

\section{\label{sec:level1}MATERIAL SYSTEM APPLICATIONS}

\subsection{Rare earth metal tritelluride}

\begin{figure}
 \includegraphics[width=0.45\textwidth]{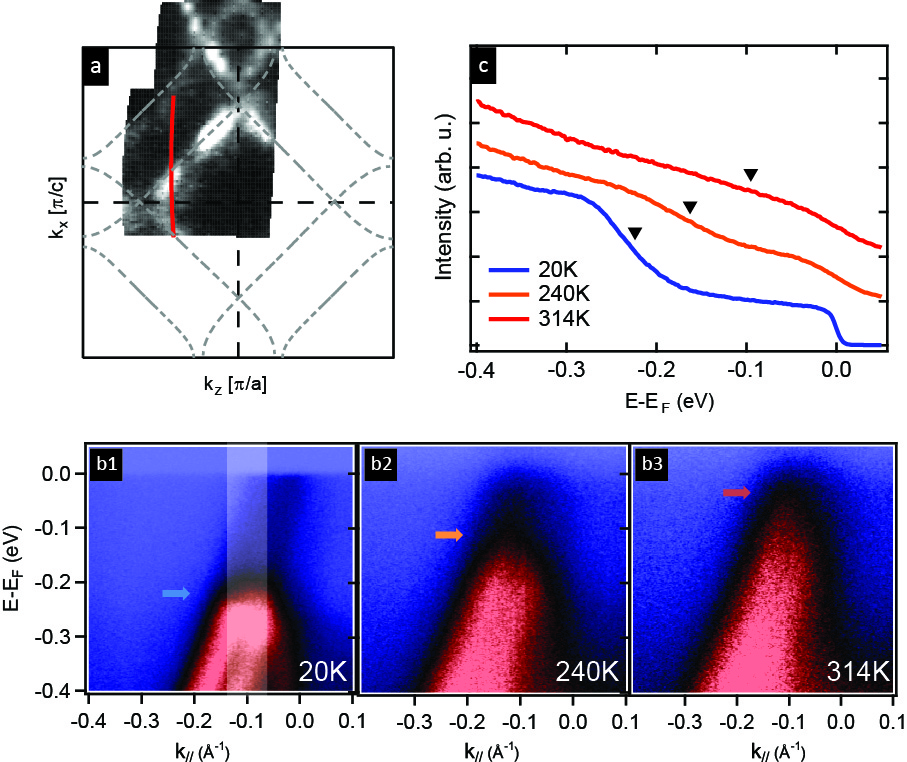}
 \caption{\label{Fig10} Measurement of TbTe$_3$ in the ac plane. (a) Fermi surface at 20K. (b1)-(b3) Temperature dependent CDW gap for the cut indicated in (a). (c) Angle integrated EDCs with integration window shaded in (b1).}
  \end{figure}

Rare earth metal tritellurides have been a model system to study charge density wave (CDW) formation, where a large CDW gap extends over the entire BZ.\cite{Moore2010,Brouet2004} 11~eV photons provide access to the entire first folded BZ (Fig.~\ref{Fig10}(a)). When the system is warmed up from 20K to 314K, the CDW order weakens. This manifests as a shrinking CDW gap below $E_F$ (b1-b3), which is observed in our experiment as an uplifting gap edge in the integrated EDC (c).

\subsection{High temperature cuprate superconductor}

\begin{figure*}[t]

 \includegraphics[width=0.9\textwidth]{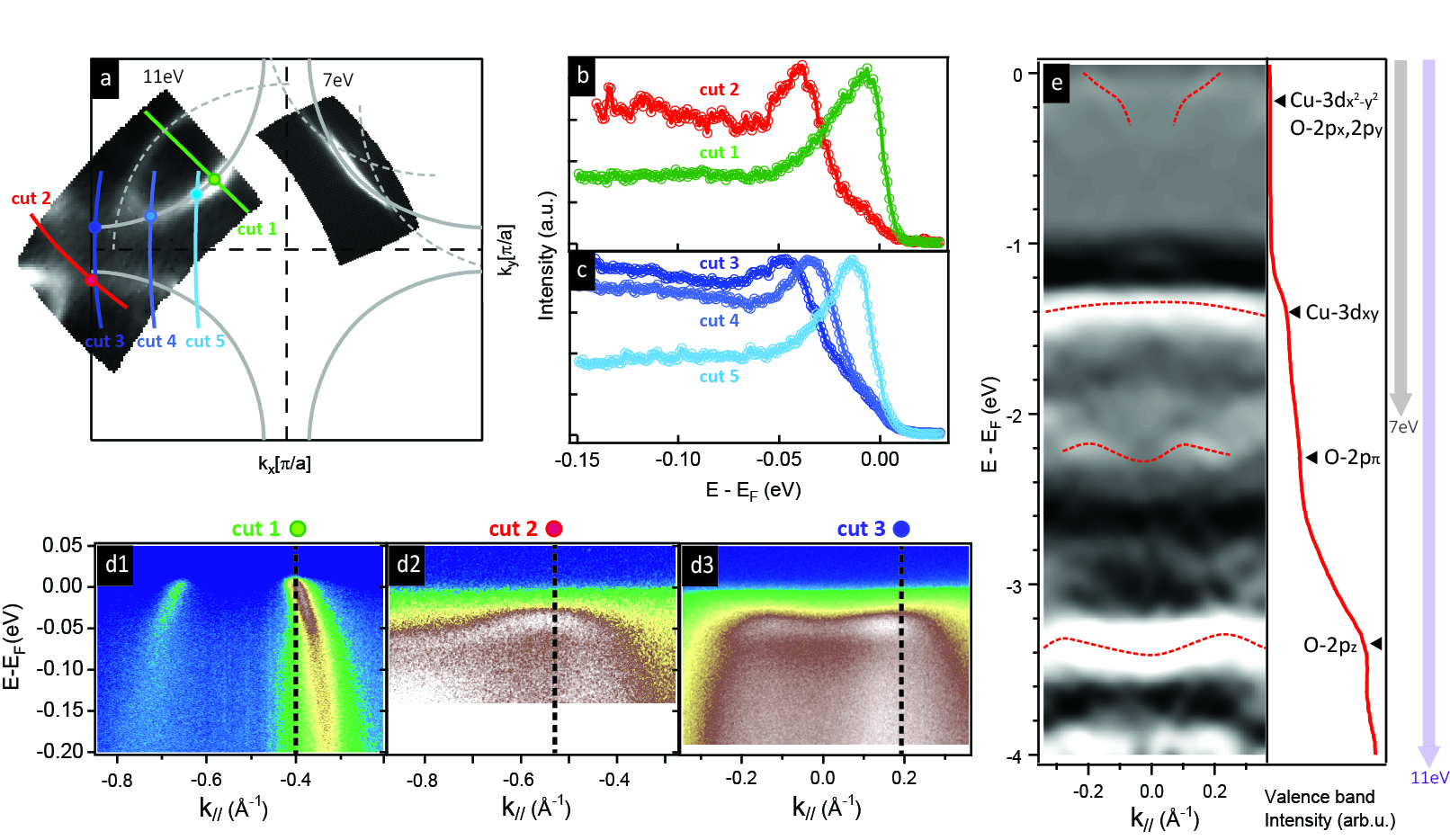}
 \caption{\label{Fig8} Measurement of overdoped Pb-Bi2212 single crystal (Tc = 80K) at 20K. (a) Comparison of Fermi surface between 11eV (left) and 7eV (right) laser. The grey dashed lines are superstructures associated with BiO layer lattice distortion. (b) Nodal (green) and antinodal (red) EDCs at Fermi momentum $k_F$, with analyzer slit parallel to zone diagonal direction. (c) EDCs at intermediate Fermi momenta between node and antinode, with the analyzer slit parallel to zone boundary. (d1)-(d3) False color plots of energy-momentum cuts from node to antinode. (e) Second derivative valence band cut in OP96 Bi2212 at the zone center. The integrated raw spectral intensity is shown as the red solid line to the right. Red dotted lines are guides to the eye.}
  \end{figure*}

We then apply the 11~eV laser-ARPES to cuprate high temperature superconductors, namely bilayer Pb-doped Bi$_2$Sr$_2$CaCu$_2$O$_{8+\delta}$ with a T$_c$ of 80K (overdoped). Previous laser-based ARPES studies at both 6~eV and 7~eV photon excitation energies have provided tremendous insights towards near-nodal excitation, yet the antinodal region near the BZ boundary was not accessible with these low photon energies.\cite{Peng2013,Parham2013,Kondo2011,Vishik2012} Fig.~\ref{Fig8}(a) compares the Fermi surface in the superconducting state obtained by 11~eV (upper left quadrant) and 7~eV (upper right quadrant) photoemission. The grey dashed lines are superstructures related to Bi-O sublattice distortion.\cite{Levin1994,Shan2003}

With 11~eV photons, the entire BZ can be mapped with high energy and momentum resolution. Notably, the 11~eV system can measure an antinodal (near ($\pi$,0)) spectrum (Fig.~\ref{Fig8}(d2-d3)), whereas measurements with 7~eV laser can barely reach the antinode and have poor cross section there.\cite{Hashimoto2012} EDCs with clear quasiparticle peaks from node to antinode are shown in panel (b) and (c) with two different analyzer slit orientations to account for different matrix element effects.\cite{Makoto2008} Here we adopted photon flux of $\sim 10^{11}$ photons/sec during data collection to ensure the absence of space-charging.

Another advantage is the ability to access valence bands (Fig.~\ref{Fig8}(e)). While a 7~eV laser cannot reach the copper 3$d_{xy}$ band (grey arrow), the 11~eV system enables access all the way down in binding energy including non-interacting oxygen $2p_z$ band (purple arrow). This is a significant advancement compared to previous laser ARPES setups in addressing materials' chemical potential evolution, and it provides opportunities for the application of a full 5-band treatment in cuprate superconductors.\cite{Kyle2004,Rosch2004,Yvonne2014}

\subsection{Iron-based superconductor}

\begin{figure*}
 \includegraphics[width=0.9\textwidth]{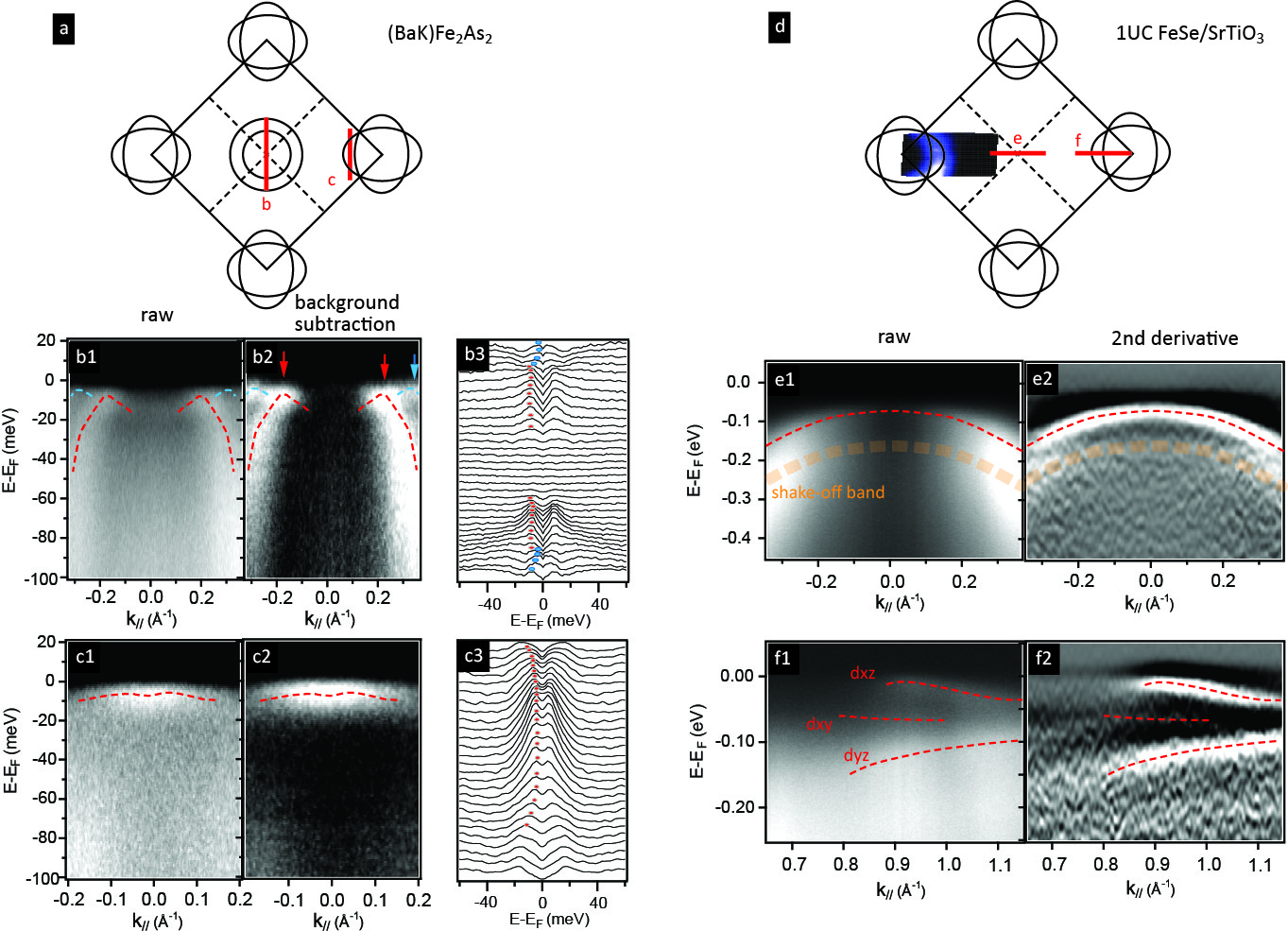}
 \caption{\label{Fig9} Measurement of iron based superconductors - single crystal (Ba,K)Fe$_2$As$_2$ (T$_c$ $\sim$ 38K, (a)-(c)) and monolayer FeSe/SrTiO$_3$ film (T$_c$ $\sim$ 65K, (d)-(f)) at T = 10K. (a)(d) Schematics of the Fermi surfaces and exemplary fermi surface map (not to scale). (b) Raw spectrum of a high symmetry cut at the $\Gamma$ in (Ba,K)Fe$_2$As$_2$ (b1), background subtracted spectrum (b2) and symmetrized EDCs (b3). The blue and red arrows indicate the gap minimum from two hole bands near $\Gamma$ point. (c) Raw spectrum of a cut near zone corner (c1), background subtracted spectrum (c2) and symmetrized EDCs (c3). The spectral quality is sufficiently good to track and fit the superconducting gap on multiple bands (blue and red dots track the quasiparticle peak dispersion). (e) Raw spectrum of a high symmetry cut at the zone center in a one unit cell FeSe/STO film (e1) and second derivative spectrum (e2). (f) Raw spectrum of a high symmetry cut at M pocket (f1) and second derivative spectrum (f2). The red lines are guide to the eye of bands with different dominating orbital components. Orange dotted lines indicate the shake-off bands due to the electron-boson coupling.}
  \end{figure*}

Previous laser-based ARPES studies on iron-based high temperature superconductors,\cite{Shimojima2010,Okazaki2012,Okazaki14092012,He2011} were limited by the small cross-section of 7~eV photons when compared to higher energy synchrotron light sources.\cite{Donghui2008,Shimojima2011} In addition, as a multi-band system, iron based superconductors have a rich Fermi surface structure, which has significant contributions from both the BZ center and corner, as the nesting-driven spin fluctuations are considered to govern the physics in these materials.\cite{Donghui2012,Zirong2013} The importance of the corner pockets has recently been highlighted in K$_x$Fe$_{2-y}$Se$_2$\cite{Zhang2011,Qian2011,Mou2011,Okazaki14092012} and the superconducting monolayer FeSe/SrTiO$_3$ film systems, which have only Fermi surfaces at the BZ corner.\cite{Liu2012,Shiyong2013,JJ2014} For ARPES studies on these systems, access to the full BZ with high resolution and flux is required.

The 11~eV laser-based ARPES system is ideal to tackle these questions. Here we demonstrate the much improved cross section and extended momentum range in both (Ba,K)Fe$_2$As$_2$ (T$_c$ $\sim$ 38K, Fig.~\ref{Fig9}(a)-(c)) bulk crystal system at 10~K and monolayer FeSe/STO film system (T$_{gap}$ $\sim$ 65K, Fig.~\ref{Fig9}(d)-(f)) at 10~K. The VUV light is polarized along the $\Gamma$-M direction, and is kept perpendicular to the cut direction.

In (Ba,K)Fe$_2$As$_2$, two hole pockets with different k$_F$\rq s centered around $\Gamma$ (Fig.~\ref{Fig9}(b))\cite{Yan2010} are identified, with the inner pocket (red arrows) showing a bigger superconducting gap (7.2~meV) than the outer pocket (blue arrows, 2.4~meV). For the two bands the Bogoliubov quasiparticle band backbending is observed in the superconducting state. For the zone corner cut (c), the d$_{xz}$/d$_{yz}$ band shows a shallow band bottom and forms an electron-like Fermi surface, consistent with previous observations from synchrotron studies.\cite{Ming2011,Shiyong2013,Ming2013}

The single unit cell FeSe film is grown on Nb-doped SrTiO$_3$ substrate, and is transferred to the ARPES measurement chamber \emph{in-situ}. The Fermi surface shows that Fermi pockets only exist at the BZ corner (Fig.~\ref{Fig9}(f)), and the hole band does not cross $E_F$ at the $\Gamma$ point (Fig.~\ref{Fig9}(e)). The superconducting backbending of the d$_{xz}$ electron band, as well as its hybridization gap with d$_{xy}$ band are observed, consistent with previous measurements at higher photon energy.\cite{Liu2012,Yan2015} The measurements also show evidence, although weaker than synchrotron data, of the shadow band due to the strong coupling to the substrate\rq s out-of-plane phonon, which could play a decisive role in boosting T$_c$ by as much as 50\% from its K$_x$Fe$_{2-y}$Se$_2$ counterpart.\cite{JJ2014} The polarization control and improved cross section of our 11~eV laser facilitates resolving gap functions with high resolution on multiple bands across the entire BZ. This is essential to understand the pairing symmetry and superconducting mechanism of this multi-band system.

%
%
%

\section{\label{sec:level1}SUMMARY}
We have presented a table-top 11eV laser-based ARPES system. The system utilizes the 9$^{th}$ harmonics of a 1024~nm IR laser to produce the 113.778~nm VUV radiation for photoemission. Combined with a Scienta R8000 hemispherical electron analyzer, we demonstrate the system\rq s energy resolution of 2~meV and momentum resolution of 0.012~\AA$^{-1}$ under realistic experimental conditions. The system is capable of reaching to $k_{\parallel}=1.2~$\AA$^{-1}$ parallel momenta as well as 5~eV in binding energy (assuming 4.5~eV material work function). Space charging effects are reduced at high repetition rates of 10~MHz. On the other hand, short pulses ($\leq$~100~ps) can be utilized for time-of-flight applications where time resolution is more prioritized.

This system provides a long desired solution to bridging the gap between the existing high-resolution small-momentum-coverage laser-based ARPES and large-momentum-coverage synchrotron-based ARPES. It is capable of producing high resolution spectra in many correlated electron systems and reaching their BZ boundaries as well as valence bands, including the temperature dependent CDW gap in rare earth metal tritelluride, the antinode and oxygen 2p band spectrum in cuprate superconductor and superconducting gap measurements in iron-pnictide and iron-chalcogenide film.

\section{\label{sec:level1}ACKNOWLEDGEMENT}
The authors acknowledge enlightening discussions with Donghui Lu and Makoto Hashimoto. Samples are kindly provided by Wei Li, Ian Fisher and Hiroshi Eisaki. A.J.M. acknowledges partial support for the VUV light source development from the National Science Foundation under SBIR grant \#0848526.  S.L.Y. acknowledges Stanford Graduate Fellowship for support. This work is a collaboration between Lumeras LLC and Stanford Institute for Materials and Energy Sciences (SIMES). The photoemission studies were supported by the Department of Energy, Office of Basic Energy Sciences, Division of Materials Sciences and Engineering.

\bibliography{GBRSIref}


\end{document}